\documentclass[12pt,epsf]{article}
\usepackage[pctex32]{graphics}
\makeatletter

\@addtoreset{equation}{section}
\newcommand{\be}{\begin{eqnarray}}
\newcommand{\ee}{\end{eqnarray}}
\newcommand{\ba}{\begin{array}}
\newcommand{\ea}{\end{array}}
\newcommand{\nn}{\nonumber}

\makeatletter \@addtoreset{equation}{section} \makeatother

\begin{document}
\vspace{1cm}
\begin{center}
~\\~\\~\\
{\bf  \LARGE Lorentz Covariant  Lagrangians of Self-dual Gauge Fields}
\vspace{1cm}

                      Wung-Hong Huang\\
                       Department of Physics\\
                       National Cheng Kung University\\
                       Tainan, Taiwan\\

\end{center}
\vspace{1cm}
\begin{center}{\bf  \Large ABSTRACT } \end{center}
We extend the method of  PST  formulation to find a systematic way  to covariantize several non-covariant Lagrangians of self-dual gauge fields.  We derive in detail the necessary basic formulas which are used to prove the existence of extra local symmetry that allows us to gauge fix the auxiliary fields therein and non-covariant formulations are restored.  We see that, the extra local symmetry  in the PST and PSST formulations, which describe the covariant Lagrangians in the 6D decomposition of $6=1+5$ and $6=3+3$ respectively, can be expressed as a simple linear form in the field strength.  However, although in this paper  we have found the covariant Lagrangians in the other decomposition of $6=2+4$,  the extra local symmetry of the gauge field cannot be expressed as a simple linear form in the field strength.  We present a no-go theorem to prove this specific property.  We also find other covariant Lagrangians with more complex decomposition of spacetime.
\vspace{3cm}
\begin{flushleft}
*E-mail:  whhwung@mail.ncku.edu.tw\\
\end{flushleft}
\newpage
\tableofcontents
\section{Introduction}
The gauge fields whose field strength is self-dual are  called chiral p-form fields.   Such fields exist only if p = 2n (n = 0, 1, . . .) and the possible spacetime dimension is D = 2(p + 1).  It is known that the chiral p-form play a central role in supergravity and in string theory and M-theory five-branes  [1].  

 Marcus and Schwarz [2] are first to see that manifest duality and spacetime covariance do not like to live in harmony with each other in Lagrangian description of chiral bosons.  Historically, the non-manifestly spacetime covariant of 0-form was proposed by Floreanini and Jackiw [3], which is then generalized to p-form by Henneaux and Teitelboim [4].  The field strength of chiral p-form $A_{1\cdot\cdot\cdot p}$ they used  splits into electric density $ {\cal E}^{i_1\cdot\cdot\cdot i_{p+1}}$ and magnetic density $ {\cal B}^{i_1\cdot\cdot\cdot i_{p+1}}$:
\be  {\cal E}_{i_1\cdot\cdot\cdot i_{p+1}} &\equiv& F_{i_1\cdot\cdot\cdot i_{p+1}}\equiv \partial_{[{i_1}}A_{i_2\cdot\cdot\cdot i_{p+1}]}\\
 {\cal B}^{i_1\cdot\cdot\cdot i_{p+1}} &\equiv &{1\over (p+1)!}\epsilon^{i_1\cdot\cdot\cdot i_{2p+2}}  F_{i_{p+2}\cdot\cdot\cdot i_{2p+2}}\equiv \tilde F^{i_1\cdot\cdot\cdot i_{p+1}} 
\ee
in which $\tilde F$ is the dual form of $F$.  The Lagrangian is described by
\be  L_{HJ}= {1\over p!}  \tilde F_{i_1\cdot\cdot\cdot i_{p+1}} ( F^{i_1\cdot\cdot\cdot i_{p+1}} - \tilde F^{i_1\cdot\cdot\cdot i_{p+1}} )
\ee
Above action lead to second class constraints and complicates the quantization procedure.  

Siegel in [5] proposed a manifestly spacetime covariant action of chiral p-form models by squaring the second-class constraints and introducing Lagrange multipliers $\lambda_{ab}$ into the action.  The Lagrangian of chiral 2 form is described by 
\be L_{Siegel}=-{1\over 12} F_{abc}F^{abc}+{1\over4}\lambda_{ab}{\cal F}^{acd}{\cal F}^b_{~cd}
\ee
in which we define
\be {\cal F}\equiv F-\tilde F
\ee
Siegel action, however, does  not have enough local symmetry to completely gauge the Lagrange multipliers away and suffers from anomaly of gauge symmetry.  

Pasti, Sorokin and Tonin in 1995 constructed a Lorentz covariant formulation of chiral p-forms in D = 2(p+1) dimensions that contains a finite number of auxiliary fields in a non-polynomial way [6,7].  For example,  6D PST Lagrangian is 
\be L_{PST} = -{1\over 6}F_{abc}F^{abc} +{1\over (\partial_q a\partial^q a)} \partial^ma(x) {\cal F}_{mnl}{\cal F}^{nlr}\partial_ra(x)
\ee
in which $a(x)$ is the auxiliary field.  In the gauge $\partial_r a =\delta_r^1$  the PST formulation reduces to the non-manifestly covariant  formulation [3,4].

Recently, a new non-covariant Lagrangian formulation of a chiral 2-form gauge field in 6D, called as (3+3) decomposition,  was derived in [8] from the Bagger-Lambert-Gustavsson (BLG) model [9]. The covariant formulation  of the associated Lagrangian  is constructed by PSST in [10], with the use of a triplet of auxiliary scalar fields. 

Later, a general non-covariant Lagrangian formulation of  self-dual gauge Theories in diverse dimensions was constructed [11].  In this general formulation the (2+4) decomposition of  Lagrangian is found.  In [12] we had also constructed a new kind of non-covariant actions of self-dual 2-form gauge theory in the decomposition of  $6=D_1+D_2+D_3$.  We also furthermore found the most  general formulation of  non-covariant  Lagrangian of self-dual gauge theory in [13].  In these paper the self-dual property of the general Lagrangian is proved in detail and it also shows that  the new non-covariant actions give field equations with 6d Lorentz invariance. 

Up to now, the PST covariant Lagrangian of self-dual gauge fields had only been constructed in the formulations of decomposition of  $6=1+5$ [6,7] and $6=3+3$ [10].  In this paper we will find a simple prescription which can  be easily extended to other non-covariant Lagrangian with more complex decomposition of spacetime.

In section 2, we first review the PST covariant Lagrangian [6,7], which essentially is to covariantize the  non-covariant Lagrangian in the decomposition of 6D spacetime into $6=1+5$.   In section 3, we covariantize the  non-covariant Lagrangian in the decomposition of spacetime into $6=2+4$ [11].  After searching the possible formulation of the associated extra local symmetry we conclude that the extra local symmetry of the two-form gauge field cannot be expressed as a simple linear form in the field strength, contrasts to the previous case.  

In section 4, we derive several useful formulas and use them to covariantize the   non-covariant Lagrangian in the decomposition of spacetime into $6=3+3$ [10].  We compare our formulation with PSST formulation and discuss the reason of why in the decomposition of   $6=2+4$ the extra local symmetry of the two-form gauge field cannot be expressed as a simple linear form in the field strength, contrasts to the PST and PSST cases. 

In section 5, we describe a simple rule from above study and argue that the method can be used to find the covariant Lagrangian associated to the generally  non-covariant self-dual gauge field [13]. As mor examples we also find the Lagrangians  in the decompositions of spacetime into $6=1+1+4$ and $6=1+2+3$ [12].  Last section is devoted to a short conclusion.  
\section{ PST Covariant Lagrangian in Decomposition:  $6=1+5$}
In the (1+5) decomposition the spacetime index $\mu= (1,\cdot\cdot\cdot,6)$ is decomposed as $\mu=(1,\dot a)$, with $\dot a=(2,\cdot\cdot\cdot,6)$. The non-covariant Lagrangian is expressed as [4]
\be L_{1+5} =- {1\over4} \tilde F_{1\dot a\dot b}(F^{1\dot a\dot b} -\tilde F^{1\dot a\dot b})
\ee

We describe the procedure of obtaining  the PST covariant Lagrangian in following three steps [6,7]. 

$\bullet$  First step: We note that 
\be
\tilde F_{1\dot a\dot b}(F^{1\dot a\dot b} -\tilde F^{1\dot a\dot b})&=&-{\cal F}_{1\dot a\dot b}{\cal F}^{1\dot a\dot b}+F_{1\dot a\dot b}(F^{1\dot a\dot b} -\tilde F^{1\dot a\dot b})\nn\\
&=&-{\cal F}_{1\dot a\dot b}{\cal F}^{1\dot a\dot b}+{1\over3}F_{\mu\nu\lambda}F^{\mu\nu\lambda}-{1\over3}F_{abc}F^{abc}-F_{1\dot a\dot b} \tilde F^{1\dot a\dot b}\nn\\
&=&-{\cal F}_{1\dot a\dot b}{\cal F}^{1\dot a\dot b}+{1\over3}F_{\mu\nu\lambda}F^{\mu\nu\lambda}+\tilde F_{1\dot a\dot b}\tilde F^{1\dot a\dot b}-F_{1\dot a\dot b} \tilde F^{1\dot a\dot b}\nn\\
&=&-{\cal F}_{1\dot a\dot b}{\cal F}^{1\dot a\dot b}+{1\over3}F_{\mu\nu\lambda}F^{\mu\nu\lambda}-\tilde F_{1\dot a\dot b}(F^{1\dot a\dot b} -\tilde F^{1\dot a\dot b})
\ee
Thus
\be F_{1\dot a\dot b}(F^{1\dot a\dot b} -\tilde F^{1\dot a\dot b})={1\over2 }\Big(-{\cal F}_{1\dot a\dot b}{\cal F}^{1\dot a\dot b}+{1\over3}F_{\mu\nu\lambda}F^{\mu\nu\lambda}\Big) 
\ee
and Lagrangian we can be expressed as  
\be L_{1+5} &=& -{1\over 24}\Big(F_{\mu\nu\lambda}F^{\mu\nu\lambda}-3{\cal F}_{1\dot a\dot b}{\cal F}^{1\dot a\dot b}\Big) 
 \ee
\\
$\bullet$ Second step : We define  two projection operators 
\be  P_\mu^{~\lambda} P_\lambda^{~\nu}=P_\mu^{~\nu},~~~~ \Pi_\mu^{~~\lambda}  \Pi_\lambda^{~\nu}= \Pi_\mu^{~\nu},~~~~P_\mu^{~~\lambda}+\Pi_\mu^{~\nu}=\delta_\mu^{~\nu}
\ee
in which $P_\mu^\nu$ is used to project direction $``1"$ while $\Pi_\mu^\nu$ is used to project direction $``\dot a"$. The projection operator $P_\mu^\nu$ is described by 
\be  P_\mu^\nu=  {\partial_\mu a ~\partial^\nu a \over (\partial a)^2}
\ee 
in which $a(r)$  is an auxiliary field.  Using above projection operator the covariant Lagrangian is expressed as
\be L_{1+5}^{PST}&=& -{1\over 24}\Big(F_{\mu\nu\lambda}F^{\mu\nu\lambda}-3{\cal F}_{\mu\nu\lambda}\cdot P^\mu_\alpha \cdot\Pi^\nu_\beta  \cdot \Pi^\lambda_\gamma \cdot {\cal F}^{\alpha\beta\gamma}\Big) \nn\\
&=& -{1\over 24}\Big(F_{\mu\nu\lambda}F^{\mu\nu\lambda}-3{\cal F}_{\mu\nu\lambda}\cdot P^\mu_\alpha\cdot (\delta ^\nu_\beta -P^\nu_\beta)\cdot (\delta ^\lambda_\gamma -P^\lambda_\gamma) \cdot {\cal F}^{\alpha\beta\gamma}\Big)\nn\\
 &=& -{1\over 24}\Big(F_{\mu\nu\lambda}F^{\mu\nu\lambda}-3{\cal F}_{\mu\nu\lambda}\cdot P^\mu_\alpha  \cdot {\cal F}^{\alpha\nu\lambda}\Big)
\ee 
as ${\cal F}_{\mu\nu\lambda}\cdot P^\mu_\alpha P^\nu_\beta~\delta ^\lambda_\gamma \cdot {\cal F}^{\alpha\beta\gamma}={\cal F}_{\mu\nu\lambda}\cdot P^\mu_\alpha P^\nu_\beta P^\lambda_\gamma \cdot {\cal F}^{\alpha\beta\gamma}=0$.\\
\\
$\bullet$ Third step : As shown in PST [6,7] there are following three useful equations
\be {\delta\over\delta A_{\alpha\beta}}\int d^6x F^{\mu\nu\lambda}F_{\mu\nu\lambda} &=&3\epsilon^{\alpha\beta\gamma\mu\nu\lambda} (\partial_\mu a) (\partial_\gamma\bar F^{(a)}_{\nu\lambda})-18\partial_\gamma\Big( P_\mu^{[\alpha }{\cal F}^{\beta\gamma]\mu}\Big)\\
{\delta\over\delta A_{\alpha\beta}}\int d^6x  {\cal F}^{\mu\nu\rho}P_\rho^\sigma {\cal F}_{\mu\nu\sigma}&=&-\epsilon^{\alpha\beta\gamma\mu\nu\lambda} (\partial_\mu a) (\partial_\gamma\bar F^{(a)}_{\nu\lambda})-6\partial_\gamma\Big( P_\mu^{[\alpha }{\cal F}^{\beta\gamma]\mu}\Big)\\
{\delta\over\delta a}\int d^6x {\cal F}^{\mu\nu\rho}P_\rho^\sigma {\cal F}_{\mu\nu\sigma}&=&2\epsilon^{\alpha\beta\gamma\mu\nu\lambda}\bar F^{(a)}_{\alpha\beta}~(\partial_\gamma\bar F^{(a)}_{\nu\lambda}) (\partial_\mu a)
\ee
in which 
\be
\bar F^{(a)}_{\mu\nu}\equiv {\cal F}_{\mu\nu\rho} {\partial^\rho a\over (\partial a)^2}
\ee
Using above three equations  the variation with respect to the associated action becomes
\be
\delta S_{1+5}^{PST}=-{1\over24}\Big[6 \epsilon^{\alpha\beta\gamma\mu\nu\lambda} (\partial_\mu a) (\partial_\gamma\bar F^{(a)}_{\nu\lambda})~\delta A_{\alpha\beta} - 6\epsilon^{\alpha\beta\gamma\mu\nu\lambda}\bar F^{(a)}_{\alpha\beta}~(\partial_\gamma\bar F^{(a)}_{\nu\lambda}) (\partial_\mu a)~\delta a
\Big]
\ee
and we have the extra local symmetry
\be
\delta a &=&\phi \\
\delta A_{\alpha\beta} &=&\phi~\bar F^{(a)}_{\alpha\beta}
\ee
in which $\phi$ is an arbitrary function.  As this Lagrangian has sufficient local symmetry it allows us to gauge fix the projection operators to become the constant matrices [7,8]
\be  P_\mu^{~\nu}=\left(\ba{cc}1&0\\0&0\ea \right),~~~~\Pi_\mu^{~\nu}=\left(\ba{cc}0&0\\0& \delta_{\dot a}^{\dot b}\ea \right),~\dot a=2,3,4,5,6.
\ee
In this gauge $L_{1+5}^{PST} = L_{1+5}$.

From  $0={\delta L_{1+5}^{PST}\over \delta A_{ab}}$ we can find the field equation of 2-form field $A_{ab}$: 
\be 0={\partial^\rho a\over \sqrt {-(\partial a)^2}} {\cal F}_{\mu\nu\rho}~~~\Rightarrow~~~{\partial^\rho a\over \sqrt {-(\partial a)^2}}{F}_{\mu\nu\rho}= {\partial^\rho a\over \sqrt {-(\partial a)^2}}{\tilde F}_{\mu\nu\rho}
\ee
Above equation is the self-duality condition in the covariant form.  Because that  after the  gauge-fixing we can get $F_{1\dot a\dot b}=\tilde F_{1\dot a\dot b}$, which is the self-duality  in the non-covariant formulation [4]. 

\section {Covariant Lagrangian in Decomposition:  $6=2+4$}
\subsection{Covariant Lagrangian}
In the (2+4) decomposition of 2-form Lagrangian in [11], the spacetime index $A$ is decomposed as $A=(a,\dot a)$, with $a=(1,2)$ and $\dot a=(3,\cdot\cdot\cdot,6)$.  The non-covariant Lagrangian is expressed as [11]
\be L_{2+4}= -{3\over 4} \Big[\tilde F_{ab\dot a}(F^{ab\dot a}-\tilde F^{ab\dot a})+ {1\over2}\tilde F_{a\dot a\dot b}(F^{a\dot a\dot b}-\tilde F^{a\dot a\dot b})\Big]
\ee
To obtain the covariant form we first note the following relation
\be  &&\tilde F_{ab\dot a}(F^{ab\dot a}-\tilde F^{ab\dot a})+{1\over2}\tilde F_{a\dot a\dot b}(F^{a\dot a\dot b}-\tilde F^{a\dot a\dot b})\nn\\
&=& -{\cal  F}_{ab\dot a}{\cal  F}^{ab\dot a}- {1\over2}{\cal  F}_{a\dot a\dot b} {\cal  F}^{a\dot a\dot b}+F_{ab\dot a}(F^{ab\dot a}-\tilde F^{ab\dot a})+{1\over2} F_{a\dot a\dot b}(F^{a\dot a\dot b}-\tilde F^{a\dot a\dot b})\nn\\
&=&-{\cal  F}_{ab\dot a}{\cal  F}^{ab\dot a}- {1\over2}{\cal  F}_{a\dot a\dot b} {\cal  F}^{a\dot a\dot b}+{1\over3}F_{\mu\nu\lambda}F^{\mu\nu\lambda}-\Big[\tilde F_{ab\dot a}(F^{ab\dot a}-\tilde F^{ab\dot a})+{1\over2}\tilde F_{a\dot a\dot b}(F^{a\dot a\dot b}-\tilde F^{a\dot a\dot b})\Big]\nn\\
\ee 
which implies
\be
\tilde F_{ab\dot a}(F^{ab\dot a}-\tilde F^{ab\dot a})+{1\over2}\tilde F_{a\dot a\dot b}(F^{a\dot a\dot b}-\tilde F^{a\dot a\dot b})={1\over 2}\Big(-{\cal  F}_{ab\dot a}{\cal  F}^{ab\dot a}- {1\over2}{\cal  F}_{a\dot a\dot b} {\cal  F}^{a\dot a\dot b}+{1\over3}F_{\mu\nu\lambda}F^{\mu\nu\lambda} \Big)\nn\\
\ee 
 and we can write $L_{2+4}$ as
\be
L_{2+4}&=&-{1\over 16}\Big(2 F_{\mu\nu\lambda}F^{\mu\nu\lambda}-6{\cal  F}_{ab\dot a}{\cal  F}^{ab\dot a}- 3{\cal  F}_{a\dot a\dot b} {\cal  F}^{a\dot a\dot b}\Big)
\ee
To find the covariant form we will  rewrite it in a more explicit form (as ``a=1,2")
\be
L_{2+4}&=&-{1\over 16}\Big(2F_{\mu\nu\lambda}F^{\mu\nu\lambda}-12{\cal  F}_{12\dot a}{\cal  F}^{12\dot a}- 3{\cal  F}_{1\dot a\dot b} {\cal  F}^{1\dot a\dot b}-3{\cal  F}_{2\dot a\dot b} {\cal  F}^{2\dot a\dot b}\Big)
\ee
We now introduce two independent projection operators 
\be 
 P_\mu^{~\alpha}&=& {\partial_\mu a \partial^\alpha a \over (\partial a)^2}\\
Q_\mu^{~\alpha}&=& {\partial_\mu b \partial^\alpha b \over (\partial b)^2}\\
\Pi_\mu^{~\alpha}&=& \delta_\mu^{~\alpha} - P_\mu^{~\alpha}- Q_\mu^{~\alpha}
\ee 
where $a$, $b$ are auxiliary fields.  Operators $P$ and $Q$ are used to project direction ``1" and ``2" respectively. 

The covariant Lagrangian we find is described by
\be
L_{2+4}^{Cov}&\equiv&-{1\over 16}\Big(2 F_{\mu\nu\lambda}F^{\mu\nu\lambda}-12{\cal  F}_{\mu\nu\lambda}\cdot P^\mu_\alpha Q^\nu_\beta \Pi^\lambda_\gamma \cdot{\cal  F}^{\alpha\beta\gamma}- 3{\cal  F}_{\mu\nu\lambda}\cdot P^\mu_\alpha \Pi^\nu_\beta \Pi^\lambda_\gamma \cdot{\cal  F}^{\alpha\beta\gamma}\nn\\
&&~~~~~~~~-3{\cal  F}_{\mu\nu\lambda}\cdot Q^\mu_\alpha \Pi^\nu_\beta \Pi^\lambda_\gamma \cdot{\cal  F}^{\alpha\beta\gamma}\Big)\nn\\
&=&-{1\over 16}\Big[ \Big(F_{\mu\nu\lambda}F^{\mu\nu\lambda}- 3{\cal  F}_{\mu\nu\lambda}\cdot P^\lambda_\gamma \cdot{\cal  F}^{\mu\nu\gamma}\Big) +\Big(F_{\mu\nu\lambda}F^{\mu\nu\lambda}- 3{\cal  F}_{\mu\nu\lambda}\cdot Q^\lambda_\gamma \cdot{\cal  F}^{\mu\nu\gamma}\Big)\Big]\nn\\
\ee
Surprisingly, above relation looks like as two kind of that in decomposition $6=1+5$.  Thus, the variation with respect to the associated action gives 
\be
\delta S_{2+4}^{Cov}=-{1\over16}\Big[6 \epsilon^{\alpha\beta\gamma\mu\nu\lambda} (\partial_\mu a) (\partial_\gamma\bar F^{(a)}_{\nu\lambda})~\delta A_{\alpha\beta} - 6\epsilon^{\alpha\beta\gamma\mu\nu\lambda}\bar F^{(a)}_{\alpha\beta}~(\partial_\gamma\bar F^{(a)}_{\nu\lambda}) (\partial_\mu a)~\delta a\nn\\
~~~~~~~~+6 \epsilon^{\alpha\beta\gamma\mu\nu\lambda} (\partial_\mu b) (\partial_\gamma\bar F^{(b)}_{\nu\lambda})~\delta A_{\alpha\beta} - 6\epsilon^{\alpha\beta\gamma\mu\nu\lambda}\bar F^{(b)}_{\alpha\beta}~(\partial_\gamma\bar F^{(b)}_{\nu\lambda}) (\partial_\mu b)~\delta b
\Big]
\ee
in which
\be
\bar F^{(a)}_{\mu\nu}&\equiv& {\cal F}_{\mu\nu\rho} {\partial^\rho a\over (\partial a)^2},~~~~~~~F^{(b)}_{\mu\nu}\equiv {\cal F}_{\mu\nu\rho} {\partial^\rho b\over (\partial b)^2}
\ee
Now, we go to final step. If we consider a local symmetry
\be
\delta a =\phi,~~~~\delta b =\chi
\ee
in which $\phi$ and $\chi$ are arbitrary functions, then the condition of $\delta L_{2+4}^{PST}=0$ is
\be
&&\sum_{\alpha\beta}\delta A_{\alpha\beta}\Big[\sum_{\gamma\mu\nu\lambda}~ \epsilon^{\alpha\beta\gamma\mu\nu\lambda} [(\partial_\mu a) (\partial_\gamma\bar F^{(a)}_{\nu\lambda})+  (\partial_\mu b) ~(\partial_\gamma\bar F^{(b)}_{\nu\lambda}) ]\Big] \nn\\
&& =\sum_{\alpha\beta}\Big[\sum_{\gamma\mu\nu\lambda}\epsilon^{\alpha\beta\gamma\mu\nu\lambda}[\bar F^{(a)}_{\alpha\beta}~(\partial_\gamma\bar F^{(a)}_{\nu\lambda}) (\partial_\mu a)~\phi+\bar F^{(b)}_{\alpha\beta}~(\partial_\gamma\bar F^{(b)}_{\nu\lambda}) (\partial_\mu b)~\chi]\Big]
\ee
Our remaining work is to find the solution of $\delta A_{\alpha\beta}$  in above equation.   The solution corresponds to the exist of the extra local symmetry which enables us to restore it to the original  non-covariant Lagrangian, as that in 6=1+5 case. 
\subsection{No-Go Theorem}
In this subsection we will search the possible formulation of the solution.  From eq.(3.13) we see that $\delta A_{\alpha\beta}$ shall be linear in the field strength ${\cal F}_{\mu\nu\rho}$.  But this does not guarantee that $\delta A_{\alpha\beta}$ is  a simple linear function of  the field strength ${\cal F}_{\mu\nu\rho}$.  For example, it may be that $\delta A_{\alpha\beta}\sim {{\cal F}_{\alpha\mu\nu} {\cal F}_{\beta}^{\mu\nu}\over  \epsilon^{ijkmnp}\partial_i {\cal F}_{ikm}\partial_n a\partial_p b}$.  However, in below we will show that the simple linear in  ${\cal F}$ could not satisfy the equation (3.13). \\

To proceed it is useful to note that there are only 3 tensors which can be used to construct the solution : ${\cal F}_{ijk}$, $\partial_\alpha a$ and $\partial_\alpha b$. We need not  $\partial_\gamma{\cal F}_{ijk}$, $\partial_\gamma \partial_\alpha a$ nor  $\partial_\gamma\partial_\alpha b$ and more derivative terms, as they can not produce the left-hand side of (3.13).  

Now, assuming that $\delta A_{\alpha\beta}$ is linear in field strength ${\cal F}_{\mu\nu\rho}$, then there are only following three possible formulations:
\\

The first possible formulation is that the indices $\alpha$ and $\beta$ appear in ${\cal F}_{\alpha\beta\rho}$ and 
\be
\delta A_{\alpha\beta}=C_1 \phi \bar F_{\alpha\beta}^{(a)}+C_2\phi \bar F_{\alpha\beta}^{(b)}
\ee
in which $C_i$ are constructed from $(\partial a)^2$ or $(\partial b)^2$. (We  consider possible terms that could produce $\phi$ term in (3.13). That produces $\chi$ term can be investigate in a similar way.)  However, in this case
\be
&&\epsilon^{\alpha\beta\gamma\mu\nu\lambda}\delta A_{\alpha\beta}\Big[ (\partial_\mu a) (\partial_\gamma\bar F^{(a)}_{\nu\lambda})+  (\partial_\mu b) ~(\partial_\gamma\bar F^{(b)}_{\nu\lambda}) \Big] \nn\\
&&=\epsilon^{\alpha\beta\gamma\mu\nu\lambda}(C_1 \phi \bar F_{\alpha\beta}^{(a)}+C_2\phi \bar F_{\alpha\beta}^{(b)})\Big[ (\partial_\mu a) (\partial_\gamma\bar F^{(a)}_{\nu\lambda})+  (\partial_\mu b) ~(\partial_\gamma\bar F^{(b)}_{\nu\lambda}) \Big] \nn\\
&& =C_1\epsilon^{\alpha\beta\gamma\mu\nu\lambda} \phi \bar F_{\alpha\beta}^{(a)}(\partial_\mu a) (\partial_\gamma\bar F^{(a)}_{\nu\lambda})+C_1\epsilon^{\alpha\beta\gamma\mu\nu\lambda} \phi \bar F_{\alpha\beta}^{(a)}(\partial_\mu b)(\partial_\gamma\bar F^{(b)}_{\nu\lambda}) \nn\\
&&+C_2\epsilon^{\alpha\beta\gamma\mu\nu\lambda} \phi \bar F_{\alpha\beta}^{(b)}(\partial_\mu a) (\partial_\gamma\bar F^{(a)}_{\nu\lambda})+C_2\epsilon^{\alpha\beta\gamma\mu\nu\lambda} \phi \bar F_{\alpha\beta}^{(b)}(\partial_\mu b)(\partial_\gamma\bar F^{(b)}_{\nu\lambda})
\ee
and it is easy to see that no matter what the values of $C_1$ and $C_2$ are it can not produce the $\phi$ term in (3.13).
\\

The second possible formulation is that only one indices $\alpha$ appears in ${\cal F}_{\alpha i j}$ and 
\be
\delta A_{\alpha\beta}=C_1 \phi \bar F_{\alpha k}^{(a)}\partial^k b \partial_\beta a+C_2 \phi \bar F_{\alpha k}^{(a)}\partial^k b \partial_\beta b +C_3 \phi \bar F_{\alpha k}^{(b)}\partial^k a \partial_\beta a+C_4 \phi \bar F_{\alpha k}^{(a)}\partial^k a \partial_\beta b 
\ee
 However, in this case  (note that $ F_{\alpha k}^{(a)}\partial^k a=0$.)

\be
&&\epsilon^{\alpha\beta\gamma\mu\nu\lambda}\delta A_{\alpha\beta}\Big[ (\partial_\mu a) (\partial_\gamma\bar F^{(a)}_{\nu\lambda})+  (\partial_\mu b) ~(\partial_\gamma\bar F^{(b)}_{\nu\lambda}) \Big] \nn\\
&&=\epsilon^{\alpha\beta\gamma\mu\nu\lambda}(C_1 \phi \bar F_{\alpha k}^{(a)}\partial^k b \partial_\beta a+C_2 \phi \bar F_{\alpha k}^{(a)}\partial^k b \partial_\beta b +C_3 \phi \bar F_{\alpha k}^{(b)}\partial^k a \partial_\beta a+C_4 \phi \bar F_{\alpha k}^{(a)}\partial^k a \partial_\beta b 
)\nn\\
&&~~~~~~~~~\times\Big[ (\partial_\mu a) (\partial_\gamma\bar F^{(a)}_{\nu\lambda})+  (\partial_\mu b) ~(\partial_\gamma\bar F^{(b)}_{\nu\lambda}) \Big] \nn\\
&& =\epsilon^{\alpha\beta\gamma\mu\nu\lambda}(C_1 \phi \bar F_{\alpha k}^{(a)}\partial^k b \partial_\beta a+C_3 \phi \bar F_{\alpha k}^{(b)}\partial^k a \partial_\beta a
)\Big[ (\partial_\mu b) ~(\partial_\gamma\bar F^{(b)}_{\nu\lambda}) \Big]\nn\\
&&+\epsilon^{\alpha\beta\gamma\mu\nu\lambda}(C_2 \phi \bar F_{\alpha k}^{(a)}\partial^k b \partial_\beta b ++C_4 \phi \bar F_{\alpha k}^{(a)}\partial^k a \partial_\beta b 
)\Big[ (\partial_\mu a) (\partial_\gamma\bar F^{(a)}_{\nu\lambda}) \Big]
\ee
and it is easy to see that no matter what the values of $C_i$ are it can not produce the $\phi$ term in (3.13).
\\

The third possible formulation is that the indices $\alpha$ and $\beta$ do not appear in ${\cal F}_{i j k}$.  In this case we have to use $\partial^i a\partial^j a\partial^k a$,  $\partial^i a\partial^j a\partial^k b$, $\partial^i a\partial^j b\partial^k b$ or $\partial^i b\partial^j b\partial^k b$ to contract the indices $ijk$ in ${\cal F}_{i j k}$. However, because ${\cal F}_{i j k}$ is a total antisymmetry tensor such contract becomes zero. 

Thus, we conclude that the extra local symmetry of the two-form gauge field cannot be expressed as a simple linear form in the field strength, contrasts to the previous case.  
\subsection{Solution}
To find the solution of (3.13) we first rewrite it as
\be
\sum_{\alpha\beta}\delta A_{\alpha\beta} (M^{(a)\alpha\beta}+N^{(b)\alpha\beta})=\sum_{\alpha\beta} \bar F^{(a)}_{\alpha\beta}M^{(a)\alpha\beta}\phi+\sum_{\alpha\beta}\bar F^{(c)}_{\alpha\beta}N^{(b)\alpha\beta}\chi
\ee
in which
\be
M^{(a)\alpha\beta} &=& \epsilon^{\alpha\beta\gamma\mu\nu\lambda}[(\partial_\mu a) (\partial_\gamma\bar F^{(a)}_{\nu\lambda})]  \\
N^{(b)\alpha\beta} &=& \epsilon^{\alpha\beta\gamma\mu\nu\lambda}[(\partial_\mu b) ~(\partial_\gamma\bar F^{(b)}_{\nu\lambda})]
\ee
Note that we have explicitly write the summation $\sum_{\alpha\beta}$. 

Although the  solution is that after the summation in (3.18) we can let each term of indices $\alpha\beta$ in left-hand is equal to that in right-hand side, i.e.
\be
\delta A_{\alpha\beta} (M^{(a)\alpha\beta}+N^{(b)\alpha\beta})= \bar F^{(a)}_{\alpha\beta}M^{(a)\alpha\beta}\phi+\bar F^{(c)}_{\alpha\beta}N^{(b)\alpha\beta}\chi
\ee
in which we do not sum over the indices $\alpha$ and $\beta$.  In this case, it has the solution  (\rm no~summation~over~the~indices $\alpha$~and~$\beta$)\\
\be
\delta A_{\alpha\beta}= { \bar F^{(a)}_{\alpha\beta}M^{(a)\alpha\beta}\phi+\bar F^{(c)}_{\alpha\beta}N^{(b)\alpha\beta}\chi\over M^{(a)\alpha\beta}+N^{(b)\alpha\beta}},  
\ee
\\
It is easy to see that, even if if $ M^{(a)\alpha\beta}+N^{(b)\alpha\beta}=0$ we can have the regular solution 
\be
\delta A_{\alpha\beta}= F^{(a)}_{\alpha\beta}~\phi-\bar F^{(c)}_{\alpha\beta}~\chi
\ee
and the above solution is well defined.

Thus the Lagrangian $L_{2+4}^{Cov}$ has sufficient local symmetry which allows us to gauge fix the projection operators to becomes the constant matrices 
\be  P_\mu^{~\nu}&=& \rm diagonal (1,0,0,0,0,0)\\
 Q_\mu^{~\nu}&=&\rm diagonal (0,1,0,0,0,0)\\
\Pi_\mu^{~\nu}&=&\rm diagonal (0,0,1,1,1,1)
\ee
In this gauge $L_{2+4}^{Cov} = L_{2+4}$. 

From  $0={\delta L_{2+4}^{Cov}\over \delta A_{ab}}$ we can find the field equation of 2-form field $A_{ab}$: 
\be 0=\partial_\mu a ~\bar F^{(a)}_{\nu\lambda}+  \partial_\mu b~ \bar F^{(b)}_{\nu\lambda}~~~\Rightarrow~~~0={\partial_\mu a\partial^\rho a\over (\partial a)^2}{\cal F}_{\nu\lambda \rho}+{\partial_\mu b\partial^\rho b\over (\partial b)^2}{\cal F}_{\nu\lambda \rho}
\ee
Above equation is the self-duality condition in the covariant form.  Because that  after the  gauge-fixing we can get $F_{a\dot a\dot b}=\tilde F_{a\dot a\dot b}$ and $F_{ab\dot a}=\tilde F_{ab\dot a}$, which are the self-duality  in the non-covariant formulation [11]. 
 
In the next section we briefly review the PSST Covariant Lagrangian [10] in decomposition:  $6=3+3$. We will see the reason of why in the decomposition of   $6=2+4$ the extra local symmetry of the two-form gauge field cannot be expressed as a simple linear form in the field strength.

\section {PSST Covariant Lagrangian in Decomposition:  $6=3+3$}
\subsection {Basic Formulation}
In the (3+3) decomposition [8] the spacetime index $\mu$ is decomposed as $\mu=(a,\dot a)$, with $a=(1,2,3)$ and $\dot a=(4,5,6)$. The non-covariant Lagrangian can be expressed as 
\be L_{3+3} &=& -{1\over 4} \Big[\tilde F_{abc}(F^{abc}-\tilde F^{abc})+3 \tilde F_{ab\dot a}(F^{ab\dot a}-\tilde F^{ab\dot a})\Big]
\ee
To obtain the covariant form we first note the following relation
\be  &&\tilde F_{abc}(F^{abc}-\tilde F^{abc})+3 \tilde F_{ab\dot a}(F^{ab\dot a}-\tilde F^{ab\dot a})\nn\\
&=& -{\cal  F}_{abc}{\cal  F}^{abc}- 3{\cal  F}_{ab\dot a} {\cal  F}^{ab\dot a}+F_{abc}(F^{abc}-\tilde F^{abc})+3 F_{ab\dot a}(F^{ab\dot a}-\tilde F^{ab\dot a})\nn\\
&=&-{\cal  F}_{abc}{\cal  F}^{abc}- 3{\cal  F}_{ab\dot a} {\cal  F}^{ab\dot a}+F_{\mu\nu\lambda}F^{\mu\nu\lambda}-F_{\dot a\dot b\dot c}F^{\dot a\dot b\dot c}-3F_{ a\dot a\dot b}F^{a\dot a\dot b} -F^{abc}\tilde F^{abc}-3 F_{ab\dot a}\tilde F^{ab\dot a}\nn\\
&=&-{\cal  F}_{abc}{\cal  F}^{abc}- 3{\cal  F}_{ab\dot a} {\cal  F}^{ab\dot a}+F_{\mu\nu\lambda}F^{\mu\nu\lambda}+\tilde F_{abc}\tilde F^{abc}+3\tilde F_{ ab\dot a}\tilde F^{ab\dot a} -F^{abc}\tilde F^{abc}-3 F_{ab\dot a}\tilde F^{ab\dot a}\nn\\
&=&-{\cal  F}_{abc}{\cal  F}^{abc}- 3{\cal  F}_{ab\dot a} {\cal  F}^{ab\dot a}+F_{\mu\nu\lambda}F^{\mu\nu\lambda}-\Big[\tilde F_{abc}(F^{abc}- F^{abc})+3 \tilde F_{ab\dot a}(F^{ab\dot a}-\tilde F^{ab\dot a})\Big]\nn\\
\ee 
which implies
\be
\tilde F_{abc}(F^{abc}-\tilde F^{abc})+3 \tilde F_{ab\dot a}(F^{ab\dot a}-\tilde F^{ab\dot a})={1\over 2}\Big(F_{\mu\nu\lambda}F^{\mu\nu\lambda}-{\cal  F}_{abc}{\cal  F}^{abc}- 3{\cal  F}_{ab\dot a} {\cal  F}^{ab\dot a} \Big)
\ee 
 and we can write $L_{3+3}$ as
\be
L_{3+3}=-{1\over 8}\Big(F_{\mu\nu\lambda}F^{\mu\nu\lambda}-{\cal  F}_{abc}{\cal  F}^{abc}- 3{\cal  F}_{ab\dot a} {\cal  F}^{ab\dot a}\Big)
\ee
Above Lagrangian is just that used by PSST in [10] to find the covariant form.  In this paper we will adopt another method to find the covariant form.  Our method is just a straightforward extending of the original PST method and can be easily extended to study other Lagrangian. 

  We introduce three independent projection operators 
\be 
 P_\mu^{~\alpha}&=& {\partial_\mu a \partial^\alpha a \over (\partial a)^2}\\
Q_\mu^{~\alpha}&=& {\partial_\mu b \partial^\alpha b \over (\partial b)^2}\\
 R_\mu^{~\alpha}&=& {\partial_\mu c \partial^\alpha c \over (\partial c)^2}\\
\Pi_\mu^{~\alpha}&=& \delta_\mu^{~\alpha} - P_\mu^{~\alpha}- Q_\mu^{~\alpha}- R_\mu^{~\alpha}
\ee 
where $a$, $b$ and $c$  are three auxiliary fields.  The operators $P$, $Q$ and $R$ are used to project direction ``1", ``2" and ``3" respectively. 

The PSST Lagrangian can be described by
\be
L_{3+3}^{PSST}&=&-{1\over 8}\Big(F_{\mu\nu\lambda}F^{\mu\nu\lambda}-6{\cal  F}_{\mu\nu\lambda}\cdot P^\mu_\alpha Q^\nu_\beta R^\lambda_\gamma \cdot{\cal  F}^{\alpha\beta\gamma}- 6{\cal  F}_{\mu\nu\lambda}\cdot P^\mu_\alpha Q^\nu_\beta \Pi^\lambda_\gamma \cdot{\cal  F}^{\alpha\beta\gamma}\nn\\
&&-6{\cal  F}_{\mu\nu\lambda}\cdot P^\mu_\alpha R^\nu_\beta \Pi^\lambda_\gamma \cdot{\cal  F}^{\alpha\beta\gamma}- 6{\cal  F}_{\mu\nu\lambda}\cdot Q^\mu_\alpha R^\nu_\beta \Pi^\lambda_\gamma\cdot {\cal  F}^{\alpha\beta\gamma}\Big)\nn\\
&=&-{1\over 8}\Big(F_{\mu\nu\lambda}F^{\mu\nu\lambda}-6{\cal  F}_{\mu\nu\lambda}\cdot P^\nu_\beta Q^\lambda_\gamma \cdot{\cal  F}^{\mu\beta\gamma}- 6{\cal  F}_{\mu\nu\lambda}\cdot P^\nu_\beta R^\lambda_\gamma \cdot{\cal  F}^{\mu\beta\gamma}\nn\\
&&-6{\cal  F}_{\mu\nu\lambda}\cdot R^\nu_\beta Q^\lambda_\gamma \cdot{\cal  F}^{\mu\beta\gamma}+12{\cal  F}_{\mu\nu\lambda}\cdot P^\mu_\alpha Q^\nu_\beta R^\lambda_\gamma\cdot {\cal  F}^{\alpha\beta\gamma}\Big)
\ee
We now will variation above Lagrangian with respect to the $A_{\alpha\beta}$ and three auxiliary fields $a$, $b$ and $c$.
\subsection {Nine Equations}
To proceed, we note that for the decompositions $6=1+5$ and $6=2+4$ we see from $L_{1+5}^{PST}$ and $L_{2+4}^{PST}$ that there is at most only one project operator in the each term of reduced covariant Lagrangian. However, for the decomposition $6=3+3$ we see from $L_{3+3}^{PSST}$  that there are two or three project operators in the each term of reduced covariant Lagrangian. Thus we need furthermore relations. 

We first note the three basic relations (all indices in the following are on 6D) which are derive in appendix
\be
{\cal F}^{abc} &=&-{1\over2}\epsilon^{abcdef}P_d^w {\cal F}_{wef}+3P_\mu^{[a} {\cal F}^{bc]\mu}\\
{\cal F}^{abc}&=&-\epsilon^{abcdef}P_d^w Q_e^s {\cal F}_{wsf} +3P_\mu^{[a} {\cal F}^{bc]\mu}+3Q_\mu^{[a} {\cal F}^{bc]\mu}-6 P_\mu^{[a}Q_\nu^{b} {\cal F}^{c]\mu\nu}\\
{\cal F}^{abc}&=&-\epsilon^{abcdef}P_d^w Q_e^s R_f^t {\cal F}_{wst} +3P_\mu^{[a} {\cal F}^{bc]\mu}+3Q_\mu^{[a} {\cal F}^{bc]\mu}+3R_\mu^{[a} {\cal F}^{bc]\mu}\nn\\
&&-6 P_\mu^{[a}Q_\nu^{b} {\cal F}^{c]\mu\nu}-6 Q_\mu^{[a}R_\nu^{b} {\cal F}^{c]\mu\nu}-6 R_\mu^{[a}P_\nu^{b} {\cal F}^{c]\mu\nu}+6P_\mu^{[a} Q_\nu^{b}R_\lambda^{c]} {\cal F}^{\mu\nu\lambda}
\ee
To obtain above equations we have used the orthogonal condition between the different projection operator, i.e $P_a^bQ_b^c=Q_a^bR_b^c=R_a^bP_b^c=0$. 

{$\bullet ~Equations~1\sim 3$} : Using  above three basic relations we have following three equations
\be -{1\over6}{\delta\over\delta A_{ab}}\int d^6x F^{\mu\nu\lambda}F_{\mu\nu\lambda} &=&\partial_c F^{abc}=\partial_c {\cal F}^{abc}\nn\\
&=&-{1\over2}\epsilon^{abcdef}\partial_c\Big(P_d^w {\cal F}_{wef}\Big)+3\partial_c\Big(P_\mu^{[a} {\cal F}^{bc]\mu}\Big)\\
&=&-\epsilon^{abcdef}\partial_c\Big(P_d^w Q_e^s {\cal F}_{wsf}\Big)+\partial_c(3P_\mu^{[a} {\cal F}^{bc]\mu}+3Q_\mu^{[a} {\cal F}^{bc]\mu}\nn\\
&&-6 P_\mu^{[a}Q_\nu^{b} {\cal F}^{c]\mu\nu})\\
&=&-\epsilon^{abcdef}\partial_c\Big(P_d^w Q_e^s {\cal F}_{wsf}\Big)+\partial_c(3P_\mu^{[a} {\cal F}^{bc]\mu}+3Q_\mu^{[a} {\cal F}^{bc]\mu}\nn\\
&&+3R_\mu^{[a} {\cal F}^{bc]\mu}-6 P_\mu^{[a}Q_\nu^{b} {\cal F}^{c]\mu\nu}-6 Q_\mu^{[a}R_\nu^{b} {\cal F}^{c]\mu\nu}\nn\\
&&-6 R_\mu^{[a}P_\nu^{b} {\cal F}^{c]\mu\nu}+6P_\mu^{[a} Q_\nu^{b}R_\lambda^{c]} {\cal F}^{\mu\nu\lambda}
\Big)
\ee
Notice that the variation of $F^{\mu\nu\lambda}F_{\mu\nu\lambda} $ with respect  $A_{\alpha\beta}$ have three different forms.  This property plays the crucial role in the following investigation.\\
{$\bullet ~Equations~4\sim 6$} : We can also use above three basic relations to derive the following equations
\be
&&{1\over2}{\delta\over\delta a}\int d^6x {\cal F}_{\mu\nu\rho}P^\rho_\lambda{\cal F}^{\mu\nu\lambda}=-\partial_\lambda\Big[{\cal F}_{\mu\nu\rho}{\partial^\rho a\over (\partial a)^2}{\cal F}^{\mu\nu\lambda}\Big]+\partial_s\Big[{\cal F}_{\mu\nu\rho}P^\rho_\lambda{\partial^s a\over (\partial a)^2}{\cal F}^{\mu\nu\lambda}\Big]\nn\\
&=&-\partial_\lambda\Big[{\cal F}_{\mu\nu\rho}{\partial^\rho a\over (\partial a)^2}\Big(-{1\over2}\epsilon^{\mu\nu\lambda abc}P_a^\sigma {\cal F}_{\sigma bc}+3P_s^{[\mu}{\cal F}^{\nu\lambda] s}\Big)\Big]+\partial_s\Big[{\cal F}_{\mu\nu\rho}P^\rho_\lambda{\partial^s a\over (\partial a)^2}{\cal F}^{\mu\nu\lambda}\Big]\nn\\
&=&-\partial_\lambda\Big[{\cal F}_{\mu\nu\rho}{\partial^\rho a\over (\partial a)^2}\Big(-{1\over2}\epsilon^{\mu\nu\lambda abc}P_a^\sigma {\cal F}_{\sigma bc}+P_s^{\lambda}{\cal F}^{\mu\nu s}\Big)\Big]+\partial_s\Big[{\cal F}_{\mu\nu\rho}P^\rho_\lambda{\partial^s a\over (\partial a)^2}{\cal F}^{\mu\nu\lambda}\Big]\nn\\
&=&\epsilon^{\mu\nu\lambda abc}\partial_\lambda\Big({\cal F}_{\mu\nu\rho}{\partial^\rho a\over (\partial a)^2}P_a^s {\cal F}_{sbc}\Big)
\ee
in which we have used the property : ${\cal F}_{\mu\nu\rho} \partial^\mu a \partial^\nu a=0$. In the same way 
\be
&&{1\over2}{\delta\over\delta a}\int d^6x {\cal F}_{\mu\nu\rho}P^\nu_\alpha Q^\rho_\beta {\cal F}^{\mu\alpha\beta}=-\partial_\lambda\Big[{\cal F}_{\mu\nu\rho}{\partial^\nu a\over (\partial a)^2}Q^\rho_\beta {\cal F}^{\mu\lambda\beta}\Big]+\partial_s\Big[{\cal F}_{\mu\nu\rho}P^\nu_\alpha Q^\rho_\beta{\partial^s a\over (\partial a)^2}{\cal F}^{\mu\alpha\beta}\Big]\nn\\
&=&-\partial_\lambda\Big[{\cal F}_{\mu\nu\rho}{\partial^\nu a\over (\partial a)^2}Q^\rho_\beta\Big(-{1\over2}\epsilon^{\mu\lambda \beta abc}P_a^\sigma {\cal F}_{\sigma bc}+3P_s^{[\mu}{\cal F}^{\lambda\beta] s}\Big)\Big]+\partial_s\Big[{\cal F}_{\mu\nu\rho}P^\nu_\alpha Q^\rho_\beta{\partial^s a\over (\partial a)^2}{\cal F}^{\mu\alpha\beta}\Big]\nn\\
&=&-\partial_\lambda\Big[{\cal F}_{\mu\nu\rho}{\partial^\nu a\over (\partial a)^2}Q^\rho_\beta\Big(-{1\over2}\epsilon^{\mu\lambda \beta abc}P_a^\sigma {\cal F}_{\sigma bc}+P_s^{\lambda}{\cal F}^{\beta \mu s}\Big)\Big]+\partial_s\Big[{\cal F}_{\mu\nu\rho}P^\nu_\alpha Q^\rho_\beta{\partial^s a\over (\partial a)^2}{\cal F}^{\mu\alpha\beta}\Big]\nn\\
&=&\epsilon^{\mu\lambda\beta abc}\partial_\lambda\Big({\cal F}_{\mu\nu\rho}{\partial^\nu a\over (\partial a)^2}P_a^k Q_\beta^\rho{\cal F}_{kbc}\Big)
\ee
in which we have used the property : ${\cal F}_{\mu\nu\rho} \partial^\mu a \partial^\nu a=0$ and orthogonality between different projection operator : $P_a^b  Q_b^d=0$.  In the same way we can follow the above method and use the orthogonality between different projection operator to derive the another equation
\be
{1\over2}{\delta\over\delta a}\int d^6x {\cal F}^{\mu\nu\rho}P_\mu^a Q_\nu^b R_\rho^c{\cal F}_{abc}=\epsilon^{\lambda bc ijk}\partial_\lambda\Big({\cal F}_{\mu\nu\rho}{\partial^\mu a\over (\partial a)^2}P_i^t Q_b^\nu R_c^\rho {\cal F}_{tjk}\Big)
\ee
{$\bullet ~Equations~7\sim 9$} : Through the simple variation we can get following equations
\be
{\delta\over\delta A_{\alpha\beta}}\int d^6x  {\cal F}^{\mu\nu\rho}P_\rho^\sigma {\cal F}_{\mu\nu\sigma}&=&- {\delta\over\delta A_{\alpha\beta}}\int d^6x  {\cal F}^{\mu\nu\rho}P_\rho^\sigma{\tilde F}_{\mu\nu\sigma}+{\delta\over\delta A_{\alpha\beta}}\int d^6x  {\cal F}^{\mu\nu\rho}P_\rho^\sigma  F_{\mu\nu\sigma}\nn\\
&=&-\epsilon^{\alpha\beta\mu\nu\sigma \lambda}\partial_\lambda \Big({\cal F}_{\mu\nu\rho}P_\sigma^\rho\Big)-6\partial_\mu\Big(P_\rho^{[\beta}{\cal F}^{\mu\alpha]\rho}\Big)
\ee
In the same we can derive the following equations
\be
{\delta\over\delta A_{\alpha\beta}}\int d^6x  {\cal F}^{\mu\nu\rho}P_\nu^a Q_\sigma^b {\cal F}_{\mu ab}&=&-\epsilon^{\alpha\beta\mu ab\lambda}\partial_\lambda \Big({\cal F}_{\mu\nu\sigma}P_a^\nu Q_b^\sigma\Big)-6\partial_\mu\Big(P_\nu^{[\alpha}Q_\rho^\beta {\cal F}^{\mu]\nu\rho}\Big)\nn\\
\\
{\delta\over\delta A_{\alpha\beta}}\int d^6x  {\cal F}^{\mu\nu\rho}P_\mu^a Q_\nu^b R_\rho^c {\cal F}_{\mu abc}&=&-\epsilon^{\alpha\beta abc\lambda}\partial_\lambda \Big({\cal F}_{\mu\nu\rho}P_a^\mu Q_b^\nu R_c^\rho\Big)-6\partial_\lambda\Big(P_\mu^{[\lambda}Q_\nu^\alpha R_\rho^{\beta]} {\cal F}^{\mu\nu\rho}\Big)\nn\\
\ee
\subsection{Existence of Extra Gauge Symmetry}
Finally, using above nine equations the variation with respect to the associated action becomes 
\be
\delta S_{3+3}^{PSST}&=&{3\over 4}\epsilon^{\alpha\beta abc\lambda}\Big[\partial_\lambda \Big({\cal F}_{a\nu\sigma}(P_b^\nu Q_c^\sigma+Q_b^\nu R_c^\sigma+R_b^\nu P_c^\sigma)-2 {\cal F}_{\mu\nu\rho}P_a^\mu Q_b^\nu R_c^\rho\Big)\delta A_{\alpha\beta}\nn\\
&&-\partial_\beta\Big( {\cal F}_{\alpha st}P_b^{[k} Q_a^{t]}{\cal F}_{kc\lambda}~{\partial^s a\over (\partial a)^2} \Big)~\delta a -\partial_\beta\Big( {\cal F}_{\alpha st}P_b^{[k} Q_a^{t]}{\cal F}_{kc\lambda}~{\partial^s b\over (\partial b)^2}) \Big)~\delta b\nn\\
&&-\partial_\beta\Big( {\cal F}_{\alpha st}Q_b^{[k} R_a^{t]}{\cal F}_{kc\lambda}~{\partial^s b\over (\partial b)^2} \Big)~\delta b-\partial_\beta\Big( {\cal F}_{\alpha st}Q_b^{[k} R_a^{t]}{\cal F}_{kc\lambda}~{\partial^s c\over (\partial c)^2}) \Big)~\delta c\nn\\
&&-\partial_\beta\Big( {\cal F}_{\alpha st}R_b^{[k} P_a^{t]}{\cal F}_{kc\lambda}~{\partial^s c\over (\partial c)^2} \Big)~\delta c-\partial_\beta\Big( {\cal F}_{\alpha st}R_b^{[k} P_a^{t]}{\cal F}_{kc\lambda}~{\partial^s a\over (\partial a)^2}) \Big)~\delta a\nn\\
&&+\partial_\alpha\Big({\cal F}_{\mu\nu\rho}P_b^{[t }Q_\beta^\nu R_a^{\rho]} {\cal F}_{tc\lambda}~{\partial^\mu a\over (\partial a)^2}\Big)~\delta a +\partial_\alpha\Big({\cal F}_{\mu\nu\rho}P_b^{[t }Q_\beta^\nu R_a^{\rho]} {\cal F}_{tc\lambda}~\delta b~{\partial^\mu b\over (\partial b)^2}\Big)~\delta b\nn\\
&& ~\partial_\alpha\Big({\cal F}_{\mu\nu\rho}P_b^{[t }Q_\beta^\nu R_a^{\rho]} {\cal F}_{tc\lambda}~{\partial^\mu c\over (\partial c)^2}) \Big)~\delta c
\Big]
\ee
which has a desired form.

To proceed, we will mention the most important point found in the investigation of this paper. 

How we can get rid of the unwanted terms, such as $\partial_\mu\Big(P_\rho^{[\beta}{\cal F}^{\mu\alpha]\rho}\Big)$, $\partial_\mu\Big(P_\nu^{[\alpha}Q_\rho^\beta {\cal F}^{\mu]\nu\rho}\Big)$ and $\partial_\lambda\Big(P_\mu^{[\lambda}Q_\nu^\alpha R_\rho^{\beta]} {\cal F}^{\mu\nu\rho}\Big)$ and finally remain only the terms proportional the $\epsilon^{\alpha\beta abc\lambda}$ in the above equation ?  The key point is that in the $L_{3+3}^{PSST}$ there is a term $F^{\mu\nu\lambda}F_{\mu\nu\lambda}$ and, as mentioned before, there are three possible forms in  the variation with respect the $A_{\alpha\beta}$. Thus, we can adjust the ratio between the three forms and get rid of the unwanted terms to have a desired form in the above equation.\\

Now, we go to final step. If we consider the local symmetry
\be
\delta a =\phi,~~~~\delta b =\chi,~~~~\delta c =\theta
\ee
in which $\phi$, $\chi$ and $\theta$ are arbitrary functions. Then the condition of $\delta S_{3+3}^{PSST}=0$ shall give us the  solution  of $\delta A_{\alpha\beta}$. Let us now see how to find the $\delta A_{\alpha\beta}$. 

First, from the rotation symmetry in  $6=3+3$ we see that there are three possible form:
\be \delta A^{(1)}_{\alpha\beta}&=&[\phi {\partial^r a\over (\partial a)^2}+\chi {\partial^r b\over (\partial b)^2}+\theta {\partial^r c\over (\partial c)^2}\Big] {\cal F}_{r \alpha\beta}\nn\\
\delta A^{(2)}_{\alpha\beta}&=&[\phi {\partial^r a\over (\partial a)^2}+\chi {\partial^r b\over (\partial b)^2}+\theta {\partial^r c\over (\partial c)^2}\Big] {\cal F}_{r a [\alpha}\Big({\partial^a a\partial_{\beta]} a\over (\partial a)^2}+{\partial^a b\partial_{\beta]} b\over (\partial b)^2}+{\partial^a c\partial_{\beta]} c\over (\partial c)^2}\Big)\nn\\
\delta A^{(3)}_{\alpha\beta}&=&[\phi {\partial^r a\over (\partial a)^2}+\chi {\partial^r b\over (\partial b)^2}+\theta {\partial^r c\over (\partial c)^2}\Big] {\cal F}_{abr}\Big({\partial^a a\partial_{[\alpha} a\over (\partial a)^2}+{\partial^a b\partial_{[\alpha} b\over (\partial b)^2}+{\partial^a c\partial_{[\alpha} c\over (\partial c)^2}\Big)\nn\\
&&\times\Big({\partial^b a\partial_{\beta]} a\over (\partial a)^2}+{\partial^b b\partial_{\beta]} b\over (\partial b)^2}+{\partial^b c\partial_{\beta]} c\over (\partial c)^2}\Big)
\ee
Then, the solution is  
\be
\delta A_{\alpha\beta}= 2\delta A^{(1)}_{\alpha\beta}+2\delta A^{(2)}_{\alpha\beta}+2\delta A^{(3)}_{\alpha\beta}= 2\phi^r Y_{rs}^{-1}\partial^\gamma a^s {\cal F}_{ab\gamma}P_{[\alpha}^{a,PSST}\Pi_{\beta]}^{b, PSST}
\ee
In the second line we express the solution in PSST notation in which  it is defined
\be
P_\mu^{\nu, PSST} &=&\partial_\mu a^r Y_{rs}^{-1}\partial^\nu a^s,\\
\Pi_{\mu}^{\nu, PSST}&=&\delta_\mu^\nu-P_\mu^{\nu, PSST}\\
Y_{rs}&=&\partial_\rho a^r \partial^\rho a^s=\delta_{rs}\Big(\partial_\rho a \partial^\rho a +\partial_\rho b \partial^\rho b +\partial_\rho c \partial^\rho c \Big)
\ee 
Note that $a^r$ are the triplet of auxiliary scalar fields, i.e. $a^1=a$,  $a^2=b$,  $a^3=c$, and $\phi^1=\phi$, $\phi^2=\chi$, $\phi^3=\theta$. 

 Thus the Lagrangian $L_{3+3}^{PSST}$ has sufficient local symmetry which allows us to gauge fix the projection operators to becomes the constant matrices 
\be  P_\mu^{~\nu}&=&\rm diagonal (1,0,0,0,0,0)\\
 Q_\mu^{~\nu}&=&\rm diagonal (0,1,0,0,0,0)\\
R_\mu^{~\nu}&=&\rm diagonal (0,0,1,0,0,0)\\
\Pi_\mu^{~\nu}&=&\rm diagonal (0,0,0,1,1,1)
\ee
In this gauge $L_{3+3}^{PSST} = L_{3+3}$ [13].

From$0={\delta L_{3+3}^{PSST}\over \delta A_{ab}}$ we can find the field equation of 2-form field $A_{ab}$ 
\be 0={\cal F}_{a\nu\sigma}(P_b^\nu Q_c^\sigma+Q_b^\nu R_c^\sigma+R_b^\nu P_c^\sigma)-2 {\cal F}_{\mu\nu\rho}P_a^\mu Q_b^\nu R_c^\rho
\ee
This is the self-duality condition in the covariant form.  Because that after the above gauge-fixing we can get $F_{abc}=\tilde F_{abc}$ and $F_{ab\dot a}=\tilde F_{ab\dot a}$, which are the self-duality  in the non-covariant formulation [11]. \\

Note that in $6=3+3$ case  $\delta A_{\mu\nu}$ is just a simple linear in field strength ${\cal F}_{abc}$, contrasts to the case of $6=2+4$.  This is because that in the case of $6=2+4$ we have only two auxiliary fields.   But in the case of $6=3+3$, as we have three auxiliary fields the third possible formulation in proving No-Go theorem in previous section cannot be ruled out.   For example ${\cal F}_{i j k}\partial^i a\partial^j b\partial^k c \ne 0$ and we have the third possible form :$\delta A^{(3)}_{\alpha\beta}$. 
\section {Covariant Lagrangian in General Decomposition}
According to above study we have found a systematic way to covariantize the non-covariant Lagrangian, which is describe in the first subsection. In the  following subsections we also find the covariant Lagrangian with more complex decomposition of spacetime. 
\subsection {General Scheme to Covariant Lagrangian}
 First, it is known that the original non-covariant Lagrangian is expressed  in terms of function $\tilde F_{abc}{\cal F}^{abc}$ [11-13].  Therefore, the first step is to express them as  ${\cal F}_{abc}{\cal F}^{abc}$.  In this step there will also appear the term of   $F_{\mu\nu\lambda}F^{\mu\nu\lambda}$. The Lagrangian form can be easily read from the function form in the original non-covariant Lagrangian.   In the second step we have to define projection operators, $P_\mu^\nu$, to render the constrained index, say ``a" into the 6d index ``$\mu$".  In the third step we can use the nine equations derived in section 2.3.2 to show that the covariant  Lagrangian has sufficiently local symmetry which allows us to gauge fix the projection operators to become the constant matrices.  In this gauge the covariant  Lagrangian becomes that originally non-covariant Lagrangian. 

\subsection {Lagrangian in Decomposition:  $6=1+1+4$}
As a further example,  let us see how to covariantize the non-covariant Lagrangian in decomposition:  $6=1+1+4$. In this case the spacetime index $\mu$ is decomposed as $\mu=(1,2,\dot a)$ and the non-covariant Lagrangian is expressed as [12]
\be L_{1+1+4}&=& -\Big[4\tilde F_{12\dot a}(F^{12\dot a}-\tilde F^{12\dot a})+ (1+\theta)\Big(\tilde F_{1\dot a\dot b}(F^{1\dot a\dot b}-\tilde F^{1\dot a\dot b})\Big)\nn\\
&&~~~~~~+ (1-\theta)\Big(\tilde F_{2\dot a\dot b}(F^{2\dot a\dot b}-\tilde F^{2\dot a\dot b})\Big)\Big]
\ee
in which $\theta$ is an arbitrary constant. To obtain the covariant form we write $L_{1+1+4}$ as
\be
L_{1+1+4}&=&{2\over 3}F_{\mu\nu\lambda}F^{\mu\nu\lambda}-4{\cal  F}_{12\dot a}{\cal  F}^{12\dot a}-(1+\theta) {\cal  F}_{1\dot a\dot b} {\cal  F}^{1\dot a\dot b}-(1-\theta) {\cal  F}_{2\dot a\dot b} {\cal  F}^{2\dot a\dot b}\Big)
\ee
We now introduce two independent projection operators 
\be 
 P_\mu^{~\alpha}&=& {\partial_\mu a \partial^\alpha a \over (\partial a)^2},~~~~~Q_\mu^{~\alpha}= {\partial_\mu b \partial^\alpha b \over (\partial b)^2}
\ee 
where $a$, $b$ are auxiliary fields.  Operators $P$ and $Q$ are used to project direction ``1" and ``2" respectively. 

The covariant Lagrangian we find is described by
\be
L_{1+1+4}^{Cov}&\equiv&-\Big[-{2\over 3}F_{\mu\nu\lambda}F^{\mu\nu\lambda}+4{\cal  F}_{\mu\nu\lambda}\cdot P^\mu_\alpha Q^\nu_\beta \Pi^\lambda_\gamma \cdot{\cal  F}^{\alpha\beta\gamma}+(1+\theta){\cal  F}_{\mu\nu\lambda}\cdot P^\mu_\alpha \Pi^\nu_\beta \Pi^\lambda_\gamma \cdot{\cal  F}^{\alpha\beta\gamma}\nn\\
&&~~~~~~~~+(1-\theta){\cal  F}_{\mu\nu\lambda}\cdot Q^\mu_\alpha \Pi^\nu_\beta \Pi^\lambda_\gamma \cdot{\cal  F}^{\alpha\beta\gamma}\Big]\nn\\
&=&-\Big[ \Big(-{2\over 3}F_{\mu\nu\lambda}F^{\mu\nu\lambda}+ (1+\theta) {\cal  F}_{\mu\nu\lambda}\cdot P^\lambda_\gamma \cdot{\cal  F}^{\mu\nu\gamma}+(1-\theta){\cal  F}_{\mu\nu\lambda}\cdot Q^\lambda_\gamma \cdot{\cal  F}^{\mu\nu\gamma}\Big]\nn\\
&=&-\Big[ \Big((1+\theta)(-{1\over 3}F_{\mu\nu\lambda}F^{\mu\nu\lambda}+  {\cal  F}_{\mu\nu\lambda}\cdot P^\lambda_\gamma \cdot{\cal  F}^{\mu\nu\gamma})\nn\\
&&~~~~~~~+(1-\theta)(-{1\over 3}F_{\mu\nu\lambda}F^{\mu\nu\lambda}+{\cal  F}_{\mu\nu\lambda}\cdot Q^\lambda_\gamma \cdot{\cal  F}^{\mu\nu\gamma})\Big]
\ee
Above relation looks like as two of that in decomposition $6=1+5$, with scale factor $(1+\theta)$ and $(1-\theta)$ before them respectively.  The variation with respect to the associated action gives 
\be
\delta S_{1+1+4}^{Cov}=-{3\over8}(1+\theta)\Big[ \epsilon^{\alpha\beta\gamma\mu\nu\lambda} (\partial_\mu a) (\partial_\gamma\bar F^{(a)}_{\nu\lambda})~\delta A_{\alpha\beta} - \epsilon^{\alpha\beta\gamma\mu\nu\lambda}\bar F^{(a)}_{\alpha\beta}~(\partial_\gamma\bar F^{(a)}_{\nu\lambda}) (\partial_\mu a)~\delta a\Big]\nn\\
-{3\over8}(1-\theta)\Big[ \epsilon^{\alpha\beta\gamma\mu\nu\lambda} (\partial_\mu b) (\partial_\gamma\bar F^{(b)}_{\nu\lambda})~\delta A_{\alpha\beta} - \epsilon^{\alpha\beta\gamma\mu\nu\lambda}\bar F^{(b)}_{\alpha\beta}~(\partial_\gamma\bar F^{(b)}_{\nu\lambda}) (\partial_\mu b)~\delta b \Big]~
\ee
Thus, as that in the case of $6=2+4$ we can find the following  local symmetry  (\rm no~summation~over~the~indices $\alpha$~and~$\beta$)
\be
\delta a &=&\phi,~~~~~\delta b =\chi\\
\delta A_{\alpha\beta}&=& { (1+\theta) \bar F^{(a)}_{\alpha\beta}M^{(a)\alpha\beta}\phi+(1-\theta)\bar F^{(b)}_{\alpha\beta}N^{(b)\alpha\beta}\chi\over (1+\theta) M^{(a)\alpha\beta}+(1-\theta)N^{(b)\alpha\beta}},  
\ee
\\
It is easy to see that, even if  $ (1+\theta) M^{(a)\alpha\beta}+(1-\theta)N^{(b)\alpha\beta}=0$ we have the regular solution 
\be
\delta A_{\alpha\beta}= F^{(a)}_{\alpha\beta}~\phi-\bar F^{(b)}_{\alpha\beta}~\chi
\ee
and the above solution is well definition anytime.  Thus, the Lagrangian $L_{1+1+4}^{Cov}$ has sufficient local symmetry which allows us to gauge fix the projection operators to becomes the constant matrices. In this gauge $L_{1+1+4}^{Cov} = L_{1+1+4}$. 
\subsection {Lagrangian in Decomposition:  $6=1+2+3$}
Let us see how to covariantize the non-covariant Lagrangian in another 
decomposition : 6= 1+2+3.  In this case the spacetime index $A$ is decomposed as $A=(1,a ,\dot a)$, with $a=(2,3)$, $\dot a=(4,5,6)$ and non-covariant  Lagrangian is [12]
\be L_{1+2+3} &=& -[\tilde F_{1ab}(F^{1ab} -\tilde F^{1ab})+\tilde F_{a\dot a\dot b}(F^{a\dot a\dot b}-\tilde F^{a\dot a\dot b})+\tilde F_{ab\dot a}(F^{ab\dot a}-\tilde F^{ab\dot a})]\nn\\ 
\ee
To obtain the covariant form we write $L_{1+2+3}$  as
\be L_{1+2+3}= -{1\over6}F_{\mu\nu\lambda}F^{\mu\nu\lambda}+{1\over2}\Big({\cal F}_{1ab}{\cal F}^{1ab} +{\cal F}_{a\dot a\dot b}{\cal F}^{a\dot a\dot b} +{\cal F}_{ab\dot a}{\cal F}^{ab\dot a}\Big) 
\ee 
We now introduce three independent projection operators
\be 
 P_\mu^{~\alpha}&=& {\partial_\mu a \partial^\alpha a \over (\partial a)^2},~~~~Q_\mu^{~\alpha}= {\partial_\mu b \partial^\alpha b \over (\partial b)^2}\\
 R_\mu^{~\alpha}&=& {\partial_\mu c \partial^\alpha c \over (\partial c)^2},~~~~~\Pi_\mu^{~\alpha}= \delta_\mu^{~\alpha} - P_\mu^{~\alpha}- Q_\mu^{~\alpha}- R_\mu^{~\alpha}
\ee 
where $a$, $b$ and $c$  are three auxiliary fields.  As in the decomposition 6=3+3, the operators $P$, $Q$ and $R$ are used to project direction ``1", ``2" and ``3" respectively. 

The Covariant Lagrangian we find is described by
\be
L_{1+2+3}^{Cov}&=&{1\over 6}\Big(-F_{\mu\nu\lambda}F^{\mu\nu\lambda}+6{\cal  F}_{\mu\nu\lambda}\cdot P^\mu_\alpha Q^\nu_\beta R^\lambda_\gamma \cdot{\cal  F}^{\alpha\beta\gamma}+ 3{\cal  F}_{\mu\nu\lambda}\cdot (Q^\mu_\alpha+R^\mu_\alpha) \Pi^\nu_\beta \Pi^\lambda_\gamma \cdot{\cal  F}^{\alpha\beta\gamma}\nn\\
&&+6{\cal  F}_{\mu\nu\lambda}\cdot Q^\mu_\alpha R^\nu_\beta \Pi^\lambda_\gamma \cdot{\cal  F}^{\alpha\beta\gamma}\nn\\
&=&{1\over 6}\Big(-F_{\mu\nu\lambda}F^{\mu\nu\lambda}+3{\cal  F}_{\mu\nu\lambda}\cdot (Q^\lambda_\gamma +R^\lambda_\gamma) \cdot{\cal  F}^{\mu\nu\gamma}-6{\cal  F}_{\mu\nu\lambda}\cdot (P^\nu_\beta Q^\lambda_\gamma+Q^\nu_\beta R^\lambda_\gamma\nn\\
&&~~~+R^\nu_\beta P^\lambda_\gamma) \cdot{\cal  F}^{\mu\beta\gamma}
+12{\cal  F}_{\mu\nu\lambda}\cdot P^\mu_\alpha Q^\nu_\beta R^\lambda_\gamma\cdot {\cal  F}^{\alpha\beta\gamma}\Big)
\ee
Using the nine equations derived  in section 2.3.2 we can quickly find that the variation with respect to the associated action becomes
\be
\delta S_{1+2+3}^{C0v}&=&{1\over 3}\epsilon^{\alpha\beta abc\lambda}\Big[6\partial_\lambda \Big({\cal F}_{a\nu\sigma}(P_b^\nu Q_c^\sigma+Q_b^\nu R_c^\sigma+R_b^\nu P_c^\sigma)-12 {\cal F}_{\mu\nu\rho}P_a^\mu Q_b^\nu R_c^\rho \nn\\
&&- 3{\cal F}_{\mu ab}(Q_c^\mu+R_c^\mu) \Big)\delta A_{\alpha\beta}+3\partial_\beta\Big( {\cal F}_{\alpha sa}Q_b^{k}{\cal F}_{kc\lambda}\delta b~{\partial^s b\over (\partial b)^2}\Big)\nn\\
&&+3\partial_\beta\Big( {\cal F}_{\alpha sa}R_b^{k}{\cal F}_{kc\lambda}\delta c~{\partial^s c\over (\partial c)^2}\Big)\nn\\
&&-6\partial_\beta\Big( {\cal F}_{\alpha st}P_b^{[k} Q_a^{t]}{\cal F}_{kc\lambda}(\delta a~{\partial^s a\over (\partial a)^2}+\delta b~{\partial^s b\over (\partial b)^2}) \Big)\nn\\
&&-6\partial_\beta\Big( {\cal F}_{\alpha st}Q_b^{[k} R_a^{t]}{\cal F}_{kc\lambda}(\delta b~{\partial^s b\over (\partial b)^2}+\delta c~{\partial^s c\over (\partial c)^2}) \Big)\nn\\
&&-6\partial_\beta\Big( {\cal F}_{\alpha st}R_b^{[k} P_a^{t]}{\cal F}_{kc\lambda}(\delta c~{\partial^s c\over (\partial c)^2}+\delta a~{\partial^s a\over (\partial a)^2}) \Big)\nn\\
&&+12\partial_\alpha\Big({\cal F}_{\mu\nu\rho}P_b^{[t }Q_\beta^\nu R_a^{\rho]} {\cal F}_{tc\lambda}(\delta a~{\partial^\mu a\over (\partial a)^2}+\delta b~{\partial^\mu b\over (\partial b)^2}+\delta c~{\partial^\mu c\over (\partial c)^2})  \Big) 
\Big]
\ee
which has a desired form. 

Now, as before, we can find a local symmetry with $\delta a =\phi$, $\delta b =\chi$, $\delta c =\theta$, and proper form of $\delta A_{\alpha\beta}$ which  can be easily read from above equation, as in the case of 6=3+3.   The existence of extra gauge symmetries  allow us to gauge fix the auxiliary fields therein and previous non-covariant formulations are reproduced. 
\\

Finally we note that in the decomposition of spacetime into $6=1+1+4$ [12] we need two auxiliary fields.  In the decomposition of spacetime into $6=1+2+3$ we need three auxiliary fields while that in the decomposition  $6=2+2+2$ we need four auxiliary fields. In the general decomposition [13] we need five auxiliary fields.  More analysis follow the above prescription can show that the covariant  Lagrangian so obtain becomes the original non-covariant Lagrangian after using the  local symmetry therein to gauge fix the projection operators to becomes the constant matrices.  The  proof of  the existence of the local symmetry is easy with the help of the nine equations derived in section 2.3.2. 
\section {Conclusion}  
In this paper we first review the PST covariant Lagrangian [6,7], which essentially is to covariantize the  non-covariant Lagrangian in the decomposition of spacetime into $6=1+5$. Then, we follow the PST method and present a straightforward method  to covariantize the  non-covariant Lagrangian in the decomposition of spacetime into $6=2+4$ and the  BLG-motivated non-covariant Lagrangian in the decomposition of spacetime into $6=3+3$.  We have derived the basic formulas which enable us to prove the existence of the local symmetry.  Using the symmetry we can gauge fix the projection operators to becomes the constant matrices and the original non-covariant Lagrangian is restored.  We have proved a no-go theorem that in the decomposition of $6=2+4$, the extra local symmetry of the gauge field cannot be expressed as a simple linear form in the field strength, contrasts to the previous two cases.

 Our method can be used to find the covariant Lagrangian associated to the generally  non-covariant self-dual gauge field. As an example we also discuss the Lagrangian  with the decomposition of spacetime into $6=1+1+4$ [12]. It is hoped that the covariantization method of straightforwardly extending from PST formulation in this paper can be applied to general systems.
\\
\\
{\bf Acknowledgments}
The author thanks Kuo-Wei Huang for discussions in the initial stage of investigation.
\\
\begin{center}{\Large \bf APPENDIX} \end{center}
\begin{appendix}
\section{Three Basic Relations}
\be
\epsilon^{abcdef}P_d^w {\cal F}_{wef} &=&{1\over6}\epsilon^{abcdef}\epsilon_{wefijk}P_d^w {\cal F}^{ijk}=-{1\over3}\epsilon^{abcd}\epsilon_{wijk}P_d^w {\cal F}^{ijk}\nn\\
&=&-{1\over3}\delta^{[abcd]}_{[wijk]}P_d^w {\cal F}^{ijk}= -2 {\cal F}_{abc}+6P_\mu^{[a} {\cal F}^{bc]\mu}\\
\epsilon^{abcdef}P_d^w Q_e^s {\cal F}_{wsf} &=&{1\over6}\epsilon^{abcdef}\epsilon_{wsfijk}P_d^w Q_e^s {\cal F}^{ijk} \nn\\
&=&-{1\over6}\delta^{[abcde]}_{[wsijk]}P_d^w Q_e^s {\cal F}^{ijk}\nn\\
&=&- {\cal F}^{abc}+3P_\mu^{[a} {\cal F}^{bc]\mu}+3Q_\mu^{[a} {\cal F}^{bc]\mu}-6 P_\mu^{[a}Q_\nu^{b} {\cal F}^{c]\mu\nu}\\
\epsilon^{abcdef}P_d^w Q_e^s R_f^t {\cal F}_{wst} &=&{1\over6}\epsilon^{abcdef}\epsilon_{wstijk}P_d^w Q_e^s R_f^t {\cal F}^{ijk} \nn\\
&=&{1\over6}\delta^{[abcdef]}_{[wstijk]}P_d^w Q_e^s R_f^t {\cal F}^{ijk} \nn\\
&=&- {\cal F}^{abc}+3P_\mu^{[a} {\cal F}^{bc]\mu}+3Q_\mu^{[a} {\cal F}^{bc]\mu}+3R_\mu^{[a} {\cal F}^{bc]\mu}\nn\\
&&-6 P_\mu^{[a}Q_\nu^{b} {\cal F}^{c]\mu\nu}-6 Q_\mu^{[a}R_\nu^{b} {\cal F}^{c]\mu\nu}-6 R_\mu^{[a}P_\nu^{b} {\cal F}^{c]\mu\nu}\nn\\
&&+6P_\mu^{[a} Q_\nu^{b}R_\lambda^{c]} {\cal F}^{\mu\nu\lambda}
\ee
To obtain above equations we have used the orthogonal condition between the projection operator, i.e $P_a^bQ_b^c=Q_a^bR_b^c=R_a^bP_b^c=0$. 
\end{appendix}
\\
\\
\\
\begin{center} {\bf REFERENCES}\end{center}
\begin{enumerate}
\item L. Alvarez-Gaume and E. Witten, Nucl. Phys. B234 (1983) 269; C.G. Callan, J.A. Harvey, and A. Strominger, Nucl. Phys. B367, 60 (1991); E. Witten,`` Five brane effective action'', J. Geom. Phys. 22 (1997)103 [hep-th/9610234].
\item N. Marcus and J.H. Schwarz, Phys. Lett. 115B (1982) 111;\\J. H. Schwarz and A. Sen, ``Duality symmetric actions," Nucl. Phys. B 411, 35 (1994) [arXiv:hep-th/9304154].
\item R. Floreanini and R. Jackiw,``Selfdual fields as charge density solitons,'' Phys. Rev. Lett. 59 (1987) 1873.
\item M. Henneaux and C. Teitelboim, ``Dynamics of chiral (self-dual) p-forms,''  Phys. Lett. B 206 (1988) 650.
\item W. Siegel,``Manifest Lorentz invariance sometimes requires nonlinearity,'' Nucl. Phys. B238 (1984) 307.
\item P. Pasti, D. Sorokin and M. Tonin, ``Note on manifest Lorentz and general coordinate invariance in duality symmetric models,'' Phys. Lett. B 352 (1995) 59
[arXiv:hep-th/9503182];\\ P. Pasti, D. Sorokin and M. Tonin, ``Duality symmetric actions with manifest space-time symmetries,'' Phys. Rev. D 52 (1995) R4277 [arXiv:hep-th/9506109]; \\P. Pasti, D. Sorokin and M. Tonin, ``Space-time symmetries in duality symmetric models,'' [arXiv:hep-th/9509052].
\item  P. Pasti, D. P. Sorokin, M. Tonin, ``On Lorentz invariant actions for chiral p-forms,'' Phys. Rev. D 55 (1997) 6292 [arXiv:hep-th/9611100].
\item P. M. Ho, Y. Matsuo, ``M5 from M2'', JHEP 0806 (2008) 105 [arXiv: 0804.3629 [hep-th]];  \\P.M. Ho, Y. Imamura, Y. Matsuo, S. Shiba, ``M5-brane in three-form flux and multiple M2-branes,'' JHEP 0808 (2008) 014, [arXiv:0805.2898 [hep-th]].
\item  J. Bagger and N. Lambert, ``Modeling multiple M2'' Phys. Rev. D 75 (2007) 045020 [arXiv:hep-th/0611108];\\
J.A. Bagger and N. Lambert, ``Gauge Symmetry and Supersymmetry of Multiple M2-Branes,'' Phys. Rev. D 77 (2008) 065008 arXiv:0711.0955 [hep-th];\\
J. Bagger and N. Lambert, ``Comments On Multiple M2-branes,'' JHEP 0802 (2008) 105, arXiv:0712.3738 [hep-th]; \\A. Gustavsson, ''Algebraic structures on parallel M2-branes,'' Nucl. Phys. B 811 (2009) 66, arXiv:0709.1260 [hep-th].
\item P. Pasti, I. Samsonov, D. Sorokin and M. Tonin,``BLG-motivated Lagrangian formulation for the chiral two-form gauge field in D = 6 and M5-branes'' Phys. Rev. D 80 (2009) 086008 [arXiv:0907.4596 [hep-th]].
\item W.-M. Chen and P.-M. Ho,``Lagrangian Formulations of Self-dual Gauge Theories in Diverse Dimensions'' Nucl. Phys. B837 (2010) 1 [arXiv:1001.3608 [hep-th]]. 
\item Wung-Hong Huang,``Lagrangian of Self-dual Gauge Fields in Various Formulations,"  Nucl. Phys. B861 (2012) 403 [arXiv:1111.5118 [hep-th]]. 
\item Wung-Hong Huang,``General Lagrangian of Non-Covariant Self-dual Gauge Field,"  JHEP 1211 (2012) 051 [arXiv:1210.1024 [hep-th]]. 
\end{enumerate}
\end{document}